\documentclass[12pt]{article}
\usepackage{graphicx}
\usepackage{hyperref}
\usepackage{color}
\usepackage{amsmath}
\usepackage{amscd,amssymb}

\newcommand{\cE}{\mathcal{E}}

\def\cR{{\mathcal R}}
\def\cS{{\mathcal S}}

\def\cE{{\mathcal E}}

\newtheorem{remark}{Remark}[section]

\newcommand{\IR}{\mathbb{R}}

\newcommand{\beq}{\begin{eqnarray}}
\newcommand{\eeq}{\end{eqnarray}}
\numberwithin{equation}{section}

\begin{document}


\begin{center}
{\large\bf Elliptic Genera and Characteristic $q$-Series of Superconformal Field Theory}
\end{center}

\vspace{0.1in}

\begin{center}
{\large
L. Bonora $^{a,}$
\footnote{bonora@sissa.it},
A. A. Bytsenko $^{(b)}$
\footnote{aabyts@gmail.com},
M. Chaichian $^{(c)}$
\footnote{masud.chaichian@helsinki.fi},
}

\vspace{5mm}
$^{(a)}$
{\it International School for Advanced Studies (SISSA/ISAS) \\
Via Bonomea 265, 34136 Trieste and INFN, Sezione di Trieste, Italy}

\vspace{0.2cm}
$^{(b)}$
{\it
Departamento de F\'{\i}sica, Universidade Estadual de
Londrina\\ Caixa Postal 6001,
Londrina-Paran\'a, Brazil}

\vspace{0.2cm}
$^{(c)}$
{\it
Department of Physics, University of Helsinki\\
P.O. Box 64, FI-00014 Helsinki, Finland}

\end{center}


\begin{abstract}
We analyze the characteristic series, the $KO$ series and the series associated with the Witten genus, and their analytic forms as the $q$-analogs of classical special functions (in particular $q$-analog of the beta integral and the gamma function). $q$-series admit an analytic interpretation in terms of
the spectral Ruelle functions, and their relations to appropriate elliptic
modular forms can be described. We show that there is a deep correspondence between the
characteristic series of the Witten genus and $KO$ characteristic series, on one side, and the denominator identities and characters of $N=2$ superconformal algebras, and the affine Lie
(super)algebras on the other. We represent the characteristic series in the form of double series using the Hecke-Rogers modular identity.
\end{abstract}

\begin{flushleft}
\\
\vspace{0.3in}
March 2015
\end{flushleft}

\newpage

\tableofcontents


\section{Introduction}

There is a deep connection between representation theory of superconformal Lie algebras,
which relies on important vertex operator algebras, and their homologies, on one side, and
elliptic genera on the other.
This manifests itself in the fact that $q$-series elliptic genera can be expressed in a quite universal way in terms of $q$-analogs of the classical special functions, in particular the spectral Ruelle functions \cite{Bytsenko13,Bonora14}. In this paper we would like to call the reader's attention on this connection by reviewing the mathematical aspects associated with it and certain physical applications. Explaining the organization of our paper gives us simultaneously the opportunity to introduce its main characters.

Sect. \ref{Prototipes} is devoted to collecting prototypes
of elliptic genera and examples. Our aim is to show that the algebraic structure necessary to construct the
elliptic genus can be expressed (in a generic way) in terms of Ruelle (Selberg-type) spectral functions of
hyperbolic geometry.

In Sect. \ref{Genus} we consider a central concept of the paper, the characteristic series of elliptic genera, a natural
topological invariant which generalizes the classical genus.
We will establish several results from classical analysis and among them the classical limit for characteristic $q$-series of the elliptic genera considered in the paper.
We show that the characteristic series of the Witten genus and the so-called $KO$ characteristic series admit an analytic interpretation in terms of the spectral Ruelle functions. Furthermore we point out the coincidence of characteristic series as the $q$-analogs of some classical special functions, in particular $q$-analogs of the beta integral and the $q$-gamma function.

In Sect. \ref{Weyl} we concentrate on (super)denominator identities and
the characters of $N=2$ superconformal algebras (Sect. \ref{N=2}).
We represent generating functions in the form of double series using the Hecke-Rogers modular identity. The classical functional equations for the Ruelle spectral functions and the modular forms in question can be used to deduce the transformation properties of supercharacters of the $N = 2$ superalgebras and denominator identities associated with the highest weight representations of affine Lie (super)algebras (see Sects. \ref{N=2} and \ref{Affine}).

Finally, in Sect. \ref{Concl}, we briefly summarize our main conclusions.
The characteristic series and denominator identities can be converted into $q$-product expressions which can inherit the homology properties of appropriate superconformal Lie algebras.

To complete this introduction let us summarize our main results:
\begin{itemize}
\item{} The characteristic series ($KO$ series and series associated with the Witten genus) has the form of $q$-analogs of
classical special functions, in particular $q$-analogs of the beta integral and the gamma function.
These $q$-series admit an analytic interpretation in terms of the spectral Ruelle functions and helps describing their relations to
appropriate elliptic modular forms.
\item{} We show a profound correspondence between the characteristic series of the Witten genus
and $KO$ characteristic series, on one side, and the denominator identities and characters of $N=2$ superconformal algebras, and the affine Lie (super)algebras $A_1^{(1)}$ and
$\widehat {{\mathfrak g}{\mathfrak l}}(m,n)$, on the other.
\item{} We represent the characteristic series in the form of double series using the Hecke-Rogers modular identity.
The main techniques we use involve spectral functions of hyperbolic three-geometry. These machinery allows us to derive interesting
new results (or rederive important old results) in superconformal field theory with it responses to three-dimensional quantum theory.
\end{itemize}

\section{Prototypes for elliptic genera}
\label{Prototipes}

For a holomorphic vector bundle $\cE$ on $X$ and a formal variable $z$ we use the following  definitions
\begin{eqnarray}
S_q(z \cE) & = & \bigoplus_{n\geq 0} (zq)^n\mbox{Sym}^n\cE,\,\,\,\,\,\,\,\,\,\,\,\,
\Lambda_q(z\cE) = \bigoplus_{n\geq 0} (zq)^n\mbox{Alt}^n\cE,
\\
S_q \left( z {\cE} \right)^{ {\mathbb C} } & = &
S_q \left( z {\cE}\right) \otimes S_q \left(
\overline{z} \overline{ {\cE} } \right),\,\,\,\,\,\,
\Lambda_q \left( z {\cE}\right)^{ {\mathbb C} } =
\Lambda_q \left( z{\cE} \right) \otimes \Lambda_q
\left( \overline{z} \overline{ {\cE} } \right)\,.
\end{eqnarray}
These formulas have good multiplicative properties and their elements should be understood as elements of the
$K$-theory of the underlying space \cite{Sharpe09}:
\begin{eqnarray}
S_q \left( {\mathcal E} \oplus {\mathcal F} \right)  & = &
\left( S_q {\mathcal E} \right) \otimes \left( S_q {\mathcal F} \right), \,\,\,\,\,\,\,
S_q \left( {\mathcal E} \ominus {\mathcal F} \right) =
\left( S_q {\mathcal E} \right) \otimes \left(
S_q {\mathcal F} \right)^{-1}\,,
\label{ident1}
\\
\Lambda_q \left( {\mathcal E} \oplus {\mathcal F} \right) & = &
\left( \Lambda_q {\mathcal E} \right) \otimes
\left( \Lambda_q {\mathcal F} \right), \,\,\,\,\,\,\,
\Lambda_q \left( {\mathcal E} \ominus {\mathcal F} \right) =
\left( \Lambda_q {\mathcal E} \right) \otimes
\left( \Lambda_q {\mathcal F} \right)^{-1}\,.
\label{ident2}
\end{eqnarray}
In Eqs. (\ref{ident1}), (\ref{ident2}) we have used the facts that
$
\mbox{Sym}^n ({\mathcal E} \oplus {\mathcal F})  =
\bigoplus_{j=0}^n \, \mbox{Sym}^j({\mathcal E}) \otimes
\mbox{Sym}^{n-j}({\mathcal F}), \,
$
$
\mbox{Alt}^n ({\mathcal E} \oplus {\mathcal F})  =
\bigoplus_{j=0}^n \, \mbox{Alt}^j({\mathcal E}) \otimes
\mbox{Alt}^{n-j}({\mathcal F})\,.
$
In the case of a line bundle ${\mathcal L}$, we have
\begin{equation}
S_q {\mathcal L} = 1 \bigoplus_{n\geq 1} q^n {\mathcal L}^n
= (1 \ominus q {\mathcal L})^{-1} = ( \Lambda_{-q} {\mathcal L})^{-1},
\end{equation}
and therefore
$\left( S_q {\cE} \right)^{-1} = \Lambda_{-q} {\cE}$
for any vector bundle ${\cE}$, and similarly
$\left( \Lambda_q {\cE} \right)^{-1} = S_{-q} {\cE}$.

{\bf Example: the Chern character.}
For the Chern polynomial $c (TX) = \prod_{j} (1+\zeta_{j})$ the resulting Chern character is
\begin{eqnarray}
\!\!\!\!\!\!\!\!
{\rm ch}(\bigotimes_{n\geq 1} S_{q^{n}}
((TX)^{\mathbb C}))\!\! & = & \!\!
\prod_j\prod_{n= 1}^\infty[(1-q^n e^{\zeta_j})(1-q^n e^{-\zeta_j})]^{-1}
\nonumber \\
\!\!& = &\!\!
\prod_j[\cR(s= (1+z_j)(1-i\varrho(\tau)))\cdot \cR(s= (1-z_j)(1-i\varrho({\tau})))]^{-1}\!,
\label{ch}
\end{eqnarray}
where $q = \exp(2\pi i\tau)$, $\varrho(\tau) = {\rm Re}\,\tau/{\rm Im}\,\tau$ and
$\zeta_j= z_j\,{\rm log}(q)$. The Ruelle zeta-functions $\cR(s)$ is a ratio of so-called Patterson-Selberg zeta-functions (see for example \cite{Bytsenko13}); it is defined for ${\rm Re}\, s\gg1$ and can be continued to a meromorphic function
on the entire complex plane $\mathbb C$. Its value $\cR(0)$ computes the $L^2$-analytic torsion of the hyperbolic three-manifold. One has the infinite product identities \cite{Bonora11,Bytsenko13}:
\begin{eqnarray}
\prod_{n=m}^{\infty}\, \big(1- q^{\mu\, n+\varepsilon} \big)
& = & \cR\big(s = (\mu \,m + \varepsilon)\,
(1-i\varrho(\tau)) + 1-\mu\big) \ ,
\label{RU1}
\\
\prod_{n=m}^{\infty}\, \big(1+ q^{\mu\,n+\varepsilon} \big)
& = & \cR\big(s = (\mu \,m + \varepsilon)\,
(1-i\varrho(\tau)) + 1-\mu+ i\sigma(\tau)\big)\ ,
\label{RU2}
\end{eqnarray}
where $\sigma(\tau) = (2\,{\rm Im}\,\tau)^{-1}$, $\mu\in\IR$, $m\geq 1$ and $\varepsilon, \nu \in {\mathbb C}$.

\section{Elliptic genera and their characteristic $q$-series}
\label{Genus}

There are various applications of modular forms in topology and physics, the elliptic genus and its generalizations being one of them. Let us recall some examples of genera:
\begin{itemize}
\item{}\, The Todd genus. Its characteristic series is the formal power series
$
z/(1-e^{-z})= \sum_{j=0}^\infty (-1)^j(B_j/j!)z^j,
$
where the $B_j$ are Bernoulli numbers. The Todd genus of a stable almost complex manifolds is an integer.
\item{}\, The $\widehat{A}$ genus. Its characteristic series is:
$
(z/2)/{\rm sinh}\,(z/2) \equiv z/(e^{z/2}- e^{-z/2}).
$
This genus is an invariant of oriented manifolds and has the property that it assumes integral values on manifolds that admit a {\it Spin} structure.
\item{}\, The Witten genus (\cite{Witten87,Witten88}) is the genus with characteristic series
\begin{eqnarray}
\!\!\!\!\!\!\!
\frac{\zeta/2}{{\rm sinh} (\zeta/2)}\prod_{n= 1}^\infty\frac{(1-q^n)^2}{(1-e^\zeta q^n)(1- e^{-\zeta}q^n)}
& \stackrel{({\rm by}\, {\rm Eq.}\, (\ref{RU1}))}{=\!=\!=\!=\!=\!=} &
\!\!\!\frac{\pi\tau{z}}{{\rm sin} (\pi\tau{z})} \left[\frac{\cR (s= 1-i\varrho(\tau))}
{\cR(s= (1+ {z})(1-i\varrho(\tau)))}\right]
\nonumber \\
& \times &
\!\!\!\!\!\!\!\!
\left[\frac{\cR (s= 1-i\varrho(\tau))}{(\cR(s= (1-{z})(1-i\varrho(\tau)))}\right]\,,
\label{WG}
\end{eqnarray}
where $\zeta = {z}\,{\rm log}(q)$. It is an even function of the variable $z$ and defines
a cobordism invariant of oriented manifolds. When applied to manifolds that admit a {\it Spin} structure it takes values in ${\mathbb Z}[[q]]$.

\item{}\, In \cite{Witten87} a genus $\chi$ of {\it Spin} manifolds with so-called $KO$ characteristic series
\begin{equation}
\sigma(L, q)  =  (L^{1/2} - L^{-1/2})\prod_{n= 1}^\infty
\frac{(1-Lq^n)(1-L^{-1}q^n)}{(1-q^n)^2}
\label{L}
\end{equation}
has been introduced. If $\tau$ is a number in the complex upper half-plane,  $\varpi\in {\mathbb C}$,
and $L = \exp(2\pi i\tau \varpi)$, $\sigma (L, q)$ is a holomorphic function of $\varpi$,
\begin{equation}
\sigma(\varpi, q) = (q^{\varpi/2} - q^{-\varpi/2})\!
\left[\frac{\cR(s= (1+ \varpi)(1-i\varrho(\tau)))}{\cR(s = 1-i\varrho(\tau))}\right]\!
\left[\frac{\cR(s= (1- \varpi)(1-i\varrho(\tau)))}{\cR(s = 1-i\varrho(\tau))}\right],
\label{WG1}
\end{equation}
which vanishes to first order of $\varpi$ at each of the points of the lattice $2\pi i(1+ \tau){\mathbb Z}$.
\end{itemize}

Let $K[[q]]$ be a spectrum representing complex $K$-theory with coefficients extended to ${\mathbb Z}[[q]]$.
Let $\Phi: XU\rightarrow K[[q]]$ be the complex orientation, where $XU\langle 2p\rangle$ is the bordism spectrum of manifolds $X$
with complex tangent bundle and trivializations of $c_1, \ldots, c_{p-1}$, so $XU\langle 2\rangle = XU$ (see for detail \cite{AndoFG}).
This complex associates to a manifold $X$ of complex dimension $d$ the genus
$
{\rm Td}\left(\bigotimes_{n\geq 1}S_{q^n}((TX) - {\mathbb C}^d)
\bigotimes_{n\geq 1}S_{q^n}((\overline{TX}) - {\mathbb C}^d)\right).
$
In addition its $K$-theory Euler class takes the form \cite{AndoFG}
\begin{equation}
\Phi (x, q) = (1-x^{-1})\prod_{n=1}^\infty \left[\frac{(1-xq^n)(1-x^{-1}q^n)}{(1-q^n)^2}\right].
\label{W}
\end{equation}
Eq. (\ref{W}) (with relationship analogous to the relation between Todd genus and $\hat{A}$ genus)
is a version of the Witten characteristic series. As it has been explained in \cite{AndoFG}, the {\it orientation} $(\delta\Phi)^\sharp$
maps a manifold $X$ of dimension $d$ to the genus
$$
\Phi(y^{-1}, d){\rm Td} \left(
\bigotimes_{n\geq 1}S_{q^n}(TX)\bigotimes_{n\geq 1}S_{q^n}(\overline{TX})
\bigotimes_{n\geq 1}\Lambda_{-yq^n}(TX)
\bigotimes_{n\geq 1}\Lambda_{-y^{-1}q^n}(\overline{TX})\right)\,.
$$
This is one of the standard formulas for the two-variable elliptic genus.
\begin{remark}
We recall briefly some well known examples of vertex operator algebra bundles which have been
used in the literature to study the elliptic genus and the Witten genus.
If $X$ is a Riemannian manifold, then the transition functions of the complex tangent bundle
$T_{\mathbb C}X$ lie in the special orthogonal group $SO(d)$, where $d$ is the dimension of $X$.
Then $\bigotimes_{n\geq 1}S_{q^n}((TX)^{\mathbb C})$ is a $V(1)^{SO(d)}$-bundle,
where $V(1)$ is the Heisenberg vertex operator algebra of dimension $d$, with $SO(d)$ as a subgroup
of Aut$(V(1))$, and $V(1)^{SO(d)}$ is the set of $SO(d)$-invariants of $V(1)$, which is a vertex operator subalgebra of $V(1)$.
$\bigotimes_{n\geq 0} \Lambda_{q^{n+1/2}}((TX)^{\mathbb C})$ is an $L(1,0)^{SO(d)}$-bundle {\rm (}we assume that $d$ is even{\rm )} where $L(1,0)$ is the level one module for the affine algebra $D_{d/2}^{(1)}$.

If $X$ is further assumed to be a spin manifold $\cS$, then $\cS\otimes\bigotimes_{n\geq 1}\Lambda_{q^{n}}((TX)^{\mathbb C})$ is also a $L(1,0)^{SO(d)}$-bundle.
If $X$ is a spin manifold and in addition the characteristic class of bundles $c_2(X)$ is trivial, then the Witten genus of $X$ admits a $q$-expansion as a modular form for the group $SL(2, {\mathbb Z})$. Thus, the Witten genus of $X$ can be written in terms of spectral Ruelle functions of hyperbolic geometry, as a holomorphic function on the upper half plane $Re\,\tau > 0$.
\end{remark}

{\bf $q$-Analog of the beta integral.}
\label{q-Analog}
For further manipulations of $q$-series we need recalling some results from classical analysis and, in particular, the $q$-analog of
the beta integral. The beta integral and its important generalization, the integral representation of the hypergeometric function,
and their $q$-analogous are:
\begin{eqnarray}
\int_0^1 t^{p-1}(1-t)^{q-1}dt  & = & \frac{\Gamma(p)\Gamma(q)}{\Gamma(p+q)}\,\,\,\,\, ({\rm Re}\, p> 0,\,\,
{\rm Re}\, q > 0)\,,
\label{G}
\\
{}_2F_1\left[\begin{matrix}
a,\, b;\, z\cr {}\, c\, {}\end{matrix}\right]
& = & \frac{\Gamma(c)}{\Gamma(b)\Gamma(c-b)}\int_0^1 t^{b-1}(1-t)^{c-b-1}(1-zt)^{-a}dt
\nonumber \\
& = & \sum_{n=0}^\infty\frac{(a)_n(b)_n z^n}{n! (c)_n}\,,
\label{F}
\end{eqnarray}
\begin{eqnarray}
\int_0^1 t^{\beta-1} (tq; q)_{\alpha-1} d(q, t) & = & \frac{\Gamma_q(\beta)\Gamma_q(\alpha)}
{\Gamma_q(\alpha+\beta)}\,,
\\
{}_2\phi_1\left[\begin{matrix}
q^\alpha,\, q^\beta;\, q,\, z\cr {}\, {}\, q^\gamma\, {}\, {}\,\end{matrix}\right]
& = &\frac{\Gamma_q(\gamma)}{\Gamma_q(\beta)\Gamma_q(\gamma - \beta)}
\int_0^1 \frac{t^{\beta-1}(tq; q)_{\gamma-\beta-1}}{(xt; q)_\alpha} d(q, t)
\nonumber \\
& = & \sum_{n=0}^\infty\frac{(a; q)_n(b; q)_n z^n}{(q; q)_n(c; q)_n}\,.
\end{eqnarray}
Here $(a)_n = a(a+1)\cdots (a+n-1)$, $(a; q)_n = \prod_{m=0}^\infty [(1-aq^m)/(1-aq^{m+n})] =
(1-a)(1-aq)\cdots (1-aq^{n-1})$, if $n$ is a nonnegative integer. The $q$-gamma function can be
defined by
\begin{equation}
\Gamma_q(z) \stackrel{\rm def}{=} \frac{(q; q)_\infty}{(q^x; q)_\infty}(1-q)^{1-z}
\equiv\prod_{n=0}^\infty \frac{1-q^{n+1}}{1-q^{n+z}}(1-q)^{1-z},\,\,\,\,
\vert q\vert <1,
\end{equation}
and hence
\begin{equation}
\Gamma_q(1+z) = \prod_{n=0}^\infty \frac{1-q^{n+1}}{1- q^{n+z+1}}(1-q)^{-z}
= \prod_{n=1}^\infty \frac{(1-q^n)(1-q^{n+1})^z}{(1-q^{n+z})(1-q^n)^z}\,.
\end{equation}
In \cite{Gosper} it has been proved that ${\rm lim}_{q\rightarrow 1^{\!-}}\,\Gamma_q(z) = \Gamma(z)$.
Consequently ${\rm lim}_{q\rightarrow 1^{\!-}}\,\Gamma_q(1-z) = \Gamma(1-z)$,\,
 ${\rm lim}_{q\rightarrow 1^{\!-}}\,\Gamma_q(1+z) = \Gamma(1+z)$.
Thus for ${\rm ch}(\bigotimes_{n\geq 1} S_{q^{n}}((TX)^{\mathbb C}))$
in Eq. (\ref{ch}) we get
\begin{equation}
\prod_j\prod_{n=1}^\infty (1-e^{\zeta_j}q^n)(1-e^{-\zeta_j}q^n) =
\frac{\prod_{n=1}^\infty (1-q^n)^2}{\prod_j\Gamma_q(1+ z_j)\Gamma_q(1-z_j)}\,.
\end{equation}
The characteristic series (\ref{WG}) (and its inverse analog (\ref{WG1})) becomes
\begin{eqnarray}
\frac{\zeta/2}{{\rm sinh} (\zeta/2)}\prod_{n= 1}^\infty\frac{(1- q^n)^2}
{(1- e^\zeta q^n)(1- e^{-\zeta}q^n)} & = &
\frac{\pi\tau z}{{\rm sin} (\pi\tau z)}\prod_{n= 1}^\infty\frac{(1- q^n)^2}{(1- q^{n+z})(1- q^{n-z})}
\nonumber \\
& = & \frac{\pi\tau z}{{\rm sin} (\pi\tau z)}\Gamma_q(1+ z)\Gamma_q(1-z)\,,
\label{q1}
\\
\lim_{\stackrel{q\rightarrow 1^-}{}} \frac{\pi\tau z}{{\rm sin} (\pi\tau z)}
\Gamma_q(1+ z)\Gamma_q(1-z)
& = & \frac{\pi\tau z}{{\rm sin} (\pi\tau z)}\Gamma(1+ z)\Gamma(1-z)
\label{cl1}
\\
& = &
\frac{\pi\tau z}{{\rm sin} (\pi\tau z)}\cdot
\frac{\pi z}{{\rm sin} (\pi z)}\,.
\label{cl2}
\end{eqnarray}
So far we have established several results from classical analysis and among them the classical limit for the characterictic series of
the Witten genus. In particular the $q$-analog of the classical result (\ref{cl1}), (\ref{cl2})
is its important generalization (\ref{q1}).
\begin{remark}

Next let us consider the following integral {\rm \cite{Askey}}:
\begin{equation}
{\mathcal I}(a; z)   =  \int_0^\infty \frac{(1+at)(1+atq)(1+atq^2)\cdots}{\!\!\!(1+t)(1+tq)(1+tq^2)\cdots}t^{z-1}dt
=
\frac{\pi}{\sin\ (\pi z)}\prod_{n=1}^\infty
\frac{(1-q^{n-z})(1-aq^{n-1})}{(1-q^n)(1-aq^{n-z-1})}.
\end{equation}
In the case $a = q^{z+1}$ we get
\begin{eqnarray}
\!\!\!\!\!\!\!\!\!
{\mathcal I}(q^{z+1}; z)  & = & \frac{\pi}{\sin\ (\pi z)}\prod_{n= 1}^\infty
\frac{(1-q^{n+z})(1-q^{n-z})}{(1-q^n)^2}
\nonumber \\
& = &
\frac{\pi}{\sin\ (\pi z)}
\left[\frac{\cR(s = (1+z)(1-i\varrho(\tau)))}{\cR(s= 1-i\varrho(\tau))}\right]\!
\left[\frac{\cR(s = (1-z)(1-i\varrho(\tau)))}{\cR(s= 1-i\varrho(\tau))}\right]\!.
\end{eqnarray}
Then the characteristic series {\rm (\ref{WG})} becomes
\begin{equation}
\frac{\pi\tau z}{{\rm sin} (\pi\tau z)}\prod_{n= 1}^\infty\frac{(1- q^n)^2}{(1- q^{n+z})(1- q^{n-z})}
= \frac{\pi\tau z}{{\rm sin} (\pi\tau z)}\cdot
\frac{\pi z}{{\rm sin} (\pi z)} {\mathcal I} (q^{z+1}; z)^{-1}.
\label{r2}
\end{equation}
The classical limit of {\rm (\ref{r2})} coincides with {\rm (\ref{cl2})}
{\rm (}it is clear that ${\rm lim}_{q\rightarrow 1^{-}}\,{\mathcal I} (q^{z+1}; z)^{-1} = 1${\rm )}.
\end{remark}

\section{Weyl character and denominator formulas}
\label{Weyl}

One of the most important features of group representations is the modular invariance of their Ka\v{c}-Weyl character formula, which allows us to derive many new results and to unify many
important results in topology.
Let us recall that the Weyl character formula in representation theory describes the characters of irreducible representations of
compact Lie groups in terms of their highest weights.
The precise statement is:  The character of an irreducible representation $V$ of a complex
semisimple Lie algebra $\mathfrak{g}$ is given by
\begin{equation}
    \operatorname{ch}(V) = \frac{\sum_{w\in W} \varepsilon(w) e^{w(\lambda+\rho)}}{e^{\rho}\prod_{\alpha \in \Delta_{+}}(1-e^{-\alpha})}\,.
\end{equation}
Here $W$ is the Weyl group,\, $\Delta_{+}$ is the subset of the positive roots of the root system
$\Delta$, \, $\rho$ is the half sum of the positive roots, \, $\lambda$ is the highest weight of the irreducible representation $V$,
\, $\varepsilon(w)$ is the determinant of the action of $w$
on the Cartan subalgebra $\mathfrak{h} \subset \mathfrak{g}$ and this is equal to $(-1)^{\ell(w)}$, where $\ell(w)$ is the length
of the Weyl group element, defined to be the minimal number of reflections with respect to simple roots such that $w$ equals
the product of those reflections.

In the special case of the trivial one-dimensional representation the character is 1, so
the Weyl character formula becomes the Weyl denominator formula:
\begin{equation}
\sum_{w\in W} \varepsilon(w)e^{w(\rho)}
= e^{\rho}\prod_{\alpha \in \Delta_{+}}(1-e^{-\alpha}).\,
\end{equation}
For special unitary groups this is equivalent to the expression
$
\sum_{\sigma \in S_n} {\rm sgn}(\sigma) \, X_1^{\sigma(1)-1} \cdots X_n^{\sigma(n)-1}
\\
=\prod_{1\le i<j\le n} (X_j-X_i)
$
i.e. the Vandermonde determinant. The Weyl character formula also holds for integrable highest weight representations
of Ka\v{c}-Moody algebras, in which case it is known as the Weyl-Ka\v{c} character formula.

{\bf (Super)denominators.} The rational exponential functions
\begin{eqnarray}
{\mathfrak D} & = & \prod_{\alpha \in \Delta_{0 +}} (1 - e^{- \alpha})
\prod_{\alpha \in \Delta_{1 +}} (1 + e^{- \alpha})^{-1},
\label{D1}
\\
\widehat{{\mathfrak D}} & = & \prod_{\alpha \in \Delta_{0 +}} (1 - e^{- \alpha})
\prod_{\alpha \in \Delta_{1 +}} (1 - e^{- \alpha})^{-1}
\nonumber \\
& = &
\prod_{\alpha \in \overline{\Delta}_{0 +}} (1 - e^{- \alpha})
\prod_{\alpha \in \Delta_{1 +}\backslash\overline{\Delta}_{1 +}} (1 + e^{- \alpha})
\prod_{\alpha \in \overline{\Delta}_{1 +}} (1 - e^{- \alpha})^{-1}\,.
\label{D2}
\end{eqnarray}
are called the {\it Weyl denominator} and the {\it Weyl superdenominator} of ${\mathfrak g}$,
respectively \cite{KacW}. Here the set of roots $\Delta = \Delta_0\cup\Delta_1$ can be represented as a disjoint union of two subsets,
called the sets of even and odd roots respectively. We introduce also the so-called {\it affine denominator} by
\begin{equation}
{\overline{\mathfrak D}} = {\mathfrak D}\prod_{n\geqslant1}(1-q^n)^\ell
\prod_{\alpha\in \Delta_0}(1-e^\alpha q^n)\prod_{\alpha\in \Delta_1}(1+ e^\alpha q^n)^{-1}\,.
\label{D3}
\end{equation}
A denominator identity for the case of affine Lie algebras is equivalent to a Macdonald identity,
while for the simplest case of the affine Lie algebra this is the Jacobi triple product identity.

Recall that in case of affine Lie algebras,
${\mathfrak g} = {\mathfrak g}(A)$, where $A$ is of finite type. The Cartan matrix of ${\mathfrak g}$
is $A = (a_{ij})_{i, j = 1, \ldots, n}$, ($a_{ii} = 2,\, a_{ij}\leq 0,\, a_{ij} = 0$ if and only if $a_{ji} = 0$),\,$a_{00} = 2$
and $a_{0i} = -2(\theta,\alpha_i)/(\alpha_i, \alpha_i)$, $a_{i0} =
(\theta, \alpha_i)/(\theta, \theta)$, a unique maximal root $\theta \in \triangle_+$.
The formal expression for the {\it partition function} $K$ of affine Lie algebras is
{\rm \cite{Kac-Peterson}}
\begin{equation}
\prod_{\alpha\in \triangle_+}(1-e^{-\alpha})^{-{\rm mult}\,\alpha} =
\sum_{\beta\in {\mathfrak h}^*}K(\beta)e^{\beta},
\label{K}
\end{equation}
where $\mathfrak h$ is the Cartan subalgebra of $\mathfrak g$, $\triangle \subset {\mathfrak h}^*$ is the root system,
$\triangle_+$ is the set of positive roots. $K$ is defined on ${\mathfrak h}^*$.
Since $(1-e^{-\alpha})^{-1} = \sum_{n=0}^\infty e^ {- n\alpha}$, $K(\beta)$ is the number of partitions of $\beta$ into sum of positive roots, where each root is counted with multiplicity.

\subsection{Infinite series generating functions}

{\bf The Hecke-Rogers modular form identity.} It has beed recognized by E. Hecke \cite{Hecke} and L. J. Rogers \cite{Rogers} that certain
modular forms could
be represented by combinations of the following double series:
\begin{equation}
\sum_{(m,n)\in  {\Omega}}(-1)^{{\mathcal H}(m,n)}q^{{\mathcal L}(m,n) + {\mathcal Q}(m, n)}
\end{equation}
Here  ${\mathcal H},\, {\mathcal L}$ are linear forms, ${\mathcal Q}$ is an indefinite quadratic form and $\Omega$ is some subset
of ${\mathbb Z}\times {\mathbb Z}$. The deepest results on this topic has come from
\cite{Kac80,Kac-Peterson} where a number of identities in the representation theory of Ka\v{c}-Moody Lie algebras has been obtained. An interesting family of modular functions
satisfying Rogers-Ramanujan type identities for arbitrary affine root systems has been obtained in \cite{Cherednik}. Extensive work in the theory of partition identities shows that basic
hypergeometric series provide the generating functions for numerous families of partition identities.

{\bf Infinite double series.} Let us introduce the folloving functions:
\begin{eqnarray}
Q_{n, k}(a_1, a_2, a_3, a_4, a_5; q) & := & (-1)^{n+k}q^{a_1n+a_2k+a_3nk+a_4n^2+a_5k^2}\,,
\\
{\mathcal J}(a_1, a_2, a_3, a_4, a_5; q) & := &
\left(\sum_{n, k\geq 0} - \sum_{n, k< 0}\right)
Q_{n, k}(a_1, a_2, a_3, a_4, a_5; q)\,,
\end{eqnarray}
The identity
\begin{eqnarray}
\sum_{n, k = -\infty}^\infty Q_{n, k\geq\vert 2n\vert}
\left(\frac{1}{2}, \frac{1}{2}, 0, -\frac{3}{2}, \frac{1}{2}; q\right)
& = &
\sum_{n,k = -\infty;\, k\geq |2n|}^\infty (-1)^{n+k}q^{(k^2-3n^2)/2+(n+k)/2}
\nonumber \\
& = & \prod_{n= 1}^\infty (1-q^n)^2
\label{DE0}
\end{eqnarray}
was conjectured by Rogers  and has been proved in \cite{Hecke,Kac-Peterson,Bressound}. We use the Rogers approach
to present the characteristic series by combination of double series (see also \cite{Andrews}).
\begin{eqnarray}
\sum_{n, k = -\infty}^\infty
Q_{n, k\geq\vert n\vert}\left(\frac{1}{2}, \frac{1}{2}, 0, \frac{1}{2}, -\frac{3}{2}; q\right)
& \stackrel{(1<\vert q\vert^{-1})}{=\!\!=\!\!=\!\!=\!\!=\!\!=} &
\prod_{j=1}^\infty \frac{(1-q^j)^2}{(1-q^{j-1})(1-q^{j})}\,.
\label{DE1}
\end{eqnarray}
\begin{eqnarray}
{\mathcal J}\left(\frac{1}{2}, \overline{x} +\frac{1}{2}, 1, \frac{1}{2}, 0; q\right) & = &
\left(\sum_{n, k\geq 0} - \sum_{n, k< 0}\right)
Q_{n, k}\left(\frac{1}{2}, \overline{x} +\frac{1}{2}, 1, \frac{1}{2}, 0; q\right)
\nonumber \\
& = &
\prod_{n=1}^\infty \frac{(1-q^n)^2}
{(1+ xq^{n-1/2})(1+ x^{-1}q^{n-1/2})}
\nonumber \\
& = &
\left[\frac{{\mathcal R}(s = 1-i\varrho(\tau))}
{{\mathcal R}(s = (\overline{x}+1/2)(1-i\varrho(\tau))+i\sigma(\tau)}\right]
\nonumber \\
& \times &
\left[\frac{{\mathcal R}(s = 1-i\varrho(\tau))}
{{\mathcal R}(s = (1-\overline{x})(1-i\varrho(\tau))+i\sigma(\tau)}\right]\,,
\label{Id2}
\end{eqnarray}
\begin{eqnarray}
\!\!\!\!\!\!\!\!\!\!\!\!\!
{\mathcal J}\left(\frac{1}{2}, \overline{x} -\frac{1+\tau}{2\tau}+\frac{1}{2}, 1,
\frac{1}{2}, 0; q\right)
& = &
\prod_{n=1}^\infty \frac{(1-q^n)^2}
{(1- xq^{n})(1- x^{-1}q^{n})}
\nonumber \\
& = &
\left[\frac{{\mathcal R}(s = 1-i\varrho(\tau))}
{{\mathcal R}(s = (1+\overline{x})(1-i\varrho(\tau)))}\right]
\nonumber \\
& \times &
\left[\frac{{\mathcal R}(s = 1-i\varrho(\tau))}
{{\mathcal R}(s = (1-\overline{x})(1-i\varrho(\tau)))}\right],
\label{Id1}
\end{eqnarray}
where ${\rm log}(x) = \overline{x}{\rm log}(q)$.
The right-hand side of the identity (\ref{Id1}), up to the $\widehat{A}$ genus, coincides with the characteristic series (\ref{WG}).

These remarkable functions may be used for other useful identities. Indeed, formula
(\ref{D3}) for the affine superdenominator $\overline{\mathfrak D}$ gives (see for detail
Example 4.1 in {\rm \cite{KacW}):
\begin{eqnarray}
{\mathcal J}(\overline{x},\overline{y},1,0,0; q)\!\!\!\!\!\!\!\! & = & \!\!\!\!\!\!\!\!
\left(\sum_{n, k \geq 0} - \sum_{n, k < 0}\right)(-1)^{n+k}q^{\overline{x}n+\overline{y}k+nk}
\nonumber \\
& \stackrel{(\vert q\vert< \vert q^{\overline{x}}\vert,\, \vert y\vert<1)}{=\!\!=\!\!=\!\!=\!\!=\!\!=\!\!=\!\!=\!\!=\!\!=\!\!=}&
\prod_{j=1}^\infty\frac{(1-q^j)^2(1-xyq^{j-1})(1-x^{-1}y^{-1}q^{j})}
{(1+ xq^{j-1})(1+ x^{-1}q^{j})(1+ yq^{j-1})(1+y^{-1}q^{j})}
\nonumber \\
& \stackrel{{\rm (by\,\, Eqs.\,\,(\ref{RU1}),\,(\ref{RU2}))}}{=\!=\!=\!=\!=\!=\!=\!\!=\!\!=\!\!=\!\!=\!\!=}&
\left[\frac{{\mathcal R}(s = 1-i\varrho(\tau)){\mathcal R}(s = 1-i\varrho(\tau))}
{{\mathcal R}(s = \overline{x}(1-i\varrho(\tau))+i\sigma(\tau)
{\mathcal R}(s = \overline{y}(1-i\varrho(\tau))+i\sigma(\tau)}\right]
\nonumber \\
&\times &  \!\!\!\!\! \!\!\!\!\! \!\!\!\!\!\!\!\!
\left[\frac{{\mathcal R}(s = (\overline{x}+ \overline{y})(1-i\varrho(\tau))+i\sigma(\tau))}
{{\mathcal R}(s = (1-\overline{x})(1-i\varrho(\tau))+i\sigma(\tau))}\right]
\nonumber \\
&\times & \!\!\!\!\! \!\!\!\!\! \!\!\!\!\!\!\!\!
\left[\frac{{\mathcal R}(s = (1-\overline{x}- \overline{y})(1-i\varrho(\tau))+i\sigma(\tau))}
{{\mathcal R}(s = (1-\overline{y})(1-i\varrho(\tau))+i\sigma(\tau))}\right]\,,
\label{DE4}
\end{eqnarray}
where ${\rm log}(x) = \overline{x}\,{\rm log}(q)$,  ${\rm log}(y) = \overline{y}\,{\rm log}(q)$.
This identity may be viewed as the denominator identity for the $N = 4$ superconformal algebra.
In the case $\vert q\vert <\vert q^{\overline{x}}\vert < 1$ we also can get {\rm \cite{KacW}}:
\begin{equation}
{\mathcal J}(\overline{x},\overline{y},1,0,0; q)
 = \sum_{j\in {\mathbb Z}}(-x)^j(1+yq^{j})^{-1}.
\end{equation}

{\bf Spectral functions of hyperbolic geometry and modular forms.}
In many problems, there are numerous instances in which we need an infinite product representation of a generating function.
Let us introduce some well-known functions and their modular properties under the action of
$SL(2, {\mathbb Z})$.
\begin{eqnarray}
\!\!\!\!\!\!\!\!\!\!\!\!\!
\varphi_1(q) & = & q^{-\frac{1}{48}}\prod_{n=1}^\infty
(1-q^{n+\frac{1}{2}})  =  q^{-\frac{1}{48}}\,{\mathcal R}(s= 3/2(1-i\varrho(\tau))) =
\frac{\eta_D(q^{\frac{1}{2}})}{\eta_D(q)}\,,
\\
\!\!\!\!\!\!\!\!\!\!\!\!\!
\varphi_2(q) & = & q^{-\frac{1}{48}}\prod_{n=1}^\infty
(1+q^{n+\frac{1}{2}})  =  q^{-\frac{1}{48}}\,{\mathcal R}(s= 3/2(1-i\varrho(\tau))+ i\sigma(\tau))
=  \frac{\eta_D(q)^2}{\eta_D(q^{\frac{1}{2}})\eta_D(q^2)}\,,
\\
\!\!\!\!\!\!\!\!\!\!\!\!\!
\varphi_3(q) & = & q^{\frac{1}{24}}\prod_{n= 1}^\infty
(1+q^{n+1})  =  q^{\frac{1}{24}}\,{\mathcal R}(s= 2(1-i\varrho(\tau))+ i\sigma(\tau))=
\frac{\eta_D(q^2)}{\eta_D(q)}\,,
\end{eqnarray}
where
$
\eta_D(q) \equiv q^{1/24}\prod_{n\in {\mathbb Z}_+}(1-q^{n})
$
is the Dedekind $\eta$-function. The linear span of $\varphi_1(q), \,\varphi_2(q)$
and $\varphi_3(q)$ is $SL(2, {\mathbb Z})$-invariant.
Since $\varphi_1(q)\cdot \varphi_2(q)\cdot \varphi_3(q) = 1$ we get
\begin{equation}
\cR(s=3/2(1-i\varrho(\tau))\cdot \cR(s =3/2(1-i\varrho(\tau)) + i\sigma(\tau))
\cdot \cR(s = 2(1-i\varrho(\tau)) + i\sigma(\tau)) = 1\,.
\end{equation}
We also have a set of useful identities:
\begin{eqnarray}
\!\!\!\!\!\!\!\!\!\!
&&
\!\!\!\!\!\!\!
{\mathcal R}(s= (\overline{x}+b)(1-i\varrho(\tau))+i\sigma(\tau))\cdot
{\mathcal R}(s= -(1+\overline{x}+b)(1-i\varrho(\tau))+i\sigma(\tau))
\nonumber \\
\!\!\!\!\!\!\!\!\!\!
=
&&
\!\!\!\!\!\!\!
q^{-\overline{x}b-b(b+1)/2}
{\mathcal R}(s= -\overline{x}(1-i\varrho(\tau))+i\sigma(\tau))\cdot
{\mathcal R}(s= (1+\overline{x})(1-i\varrho(\tau))+i\sigma(\tau))
\nonumber \\
\!\!\!\!\!\!\!\!\!\!
=
&&
\!\!\!\!\!\!\!
q^{-\overline{x}(b-1)-b(b+1)/2}
{\mathcal R}(s= (1-\overline{x})(1-i\varrho(\tau))+i\sigma(\tau))\cdot
{\mathcal R}(s= \overline{x}(1-i\varrho(\tau))+i\sigma(\tau)).
\label{DE3}
\end{eqnarray}
The classical functional equations for the Ruelle spectral functions and the modular forms
in question can be used to deduce the transformation properties of the character modules of
$N = 2$ superalgebras.

\subsection{$N=2$ superconformal algebras}
\label{N=2}

The superconformal Lie algebras give rise to some of the most important vertex algebras. The Neveu-Schwarz
(NS) algebra is the simplest among the superconformal  Lie algebras. The next simplest after the NS algebra
is the $N=2$ superconformal Lie algebra. The latter is a graded  superalgebra spanned by a central element
$\mathcal C$, a Virasoro formal distribution $L(z)$, an even formal distribution $J(z)$ primary with respect
to $L(z)$ of conformal weight 1, and two odd primary with respect to $L(z)$ formal distributions $G^+(z)$
and $G^-(z)$ of conformal weight 3/2. The remaining operator products (OPE) can be written as follows:
\begin{eqnarray}
&&
J(z)J(w) \sim \frac{{\mathcal C}/3}{(z - w)^2}, \,\,\,\,\,\,\,
J(z)G^\pm (w) \sim \pm \frac{G^\pm (w)}{z-w}, \,\,\,\,\,\,\, G^\pm (z) G^\pm (w) \sim 0,
\nonumber \\
&&
G^+(z)G^-(w)  \sim  \frac{{\mathcal C}/3}{(z - w)^3}
+ \frac{J(w)}{(z - w)^2} + \frac{L(w)+ \partial(J(w)/2)}{z-w}.
\end{eqnarray}
This superalgebra contains the NS subalgebra spanned by ${\mathcal C}, \, L(z)$ and
$G(z) = G^+(z) + G^-(z)$ and the NS subalgebra spanned by ${\mathcal C}, \, L(z)$ and
$\overline{G}(z) = i(G^+(z) - G^-(z))$ \cite{KacW}.

In \cite{Matsuo,Dobrev,Kiritsis} the characters of the irreducible unitary representations of the $N=2$ superconformal algebras $\chi^{A, P^\pm}(x, y)$ with ${\mathcal C}< 1$ were computed. For the $A, P^\pm$ algebras they have been found to be given by
\begin{eqnarray}
\chi_{j, k}^{(m), A}(X, Y) & = & X^{h_{j, k}^A}Y^{q_{j, k}^A}\Psi_A(Y, X)\Gamma_{j, k}^{(m)}(Y, X)\,,
\\
\chi_{j, k}^{(m), P^\pm}(X, Y) & = & X^{h_{j, k}^{P^\pm}}Y^{q_{j, k}^{P^\pm}}\Psi_P(Y, X)
\Gamma_{j, k}^{(m)}(Y^{\pm 1}, X)\,,
\end{eqnarray}
where $h_{j, k}^A = (jk-1/4)/m$, $h_{j, k}^{P^\pm}= jk/m + {\mathcal C}/8$, $q_{j, k}^A= (j-k)/m$,
$q_{j, k}^{P^\pm} = \pm(j-k)/m$, $j, k \in {\mathbb Z}+1/2$, $m\in {\mathbb Z}_+- 1$ and
\begin{eqnarray}
\Psi_A(Y, X) & = & \prod_{n=1}^\infty \frac{(1+YX^{n-1/2})(1+Y^{-1}X^{n-1/2})}{(1-X^n)^2}
\nonumber \\
& \stackrel{({\rm by}\, {\rm Eq.}\, (\ref{RU2}))}{=\!=\!=\!=\!=\!=} &
\left[\frac{\cR(s = (1/2+\varepsilon)(1-i\varrho(\tau))+i\sigma(\tau))}{\cR(s= 1-i\varrho(\tau))}\right]
\nonumber \\
& \times &
\left[\frac{\cR(s = (1/2- \varepsilon)(1-i\varrho(\tau))+i\sigma(\tau))}
{\cR(s= 1-i\varrho(\tau))}\right]\,,
\label{F1}
\end{eqnarray}
\begin{eqnarray}
\Psi_P(Y, X) & = & (Y^{1/2} + Y^{-1/2})\prod_{n=1}^\infty \frac{(1+YX^n)(1+Y^{-1}X^n)}{(1-X^n)^2}
\nonumber \\
& \stackrel{({\rm by}\, {\rm Eq.}\, (\ref{RU2}))}{=\!=\!=\!=\!=\!=} &
2{\rm cos}(\pi \tau\varepsilon)\left[\frac{\cR(s = (1+ \varepsilon)(1-i\varrho(\tau))+i\sigma(\tau))}
{\cR(s= 1-i\varrho(\tau))}\right]
\nonumber \\
& \times &
\left[\frac{\cR(s = (1- \varepsilon)(1-i\varrho(\tau))+i\sigma(\tau))}
{\cR(s= 1-i\varrho(\tau))}\right]\,,
\label{F2}
\end{eqnarray}
\begin{equation}
\Gamma_{j, k}^{(m)}(Y, X) = Y\sum_{n = -\infty}^{\infty} X^{mn^2+(j+k)n}
\left(\frac{1}{Y+X^{mn+j}} + \frac{1}{Y+X^{-mn-k}}\right)\,.
\end{equation}
In Eqs. (\ref{F1}) and (\ref{F2}) $X=\exp(2\pi i\tau)\equiv q,\, {\rm log}(Y)=
\varepsilon{\rm log}(q)$. There is a profound correspondence between the characteristic series for
the elliptic genus and the characters (and denominator identities) for $N=2$ superconformal
algebras $A, P^\pm$:
\begin{eqnarray}
{\mathcal J}\left(\frac{1}{2},\overline{x}+\frac{1}{2}, \frac{1}{2}, 1, 0; q\right)
\!\!\!\!\!\!\!\!\!\! & = & \!\!\!\!\!\!\!\!\!\! \left(\sum_{n, k\geq 0} - \sum_{n, k< 0} \right) q^{\frac{1}{2}n(n+1)
+ (n+\frac{1}{2})k + \overline{x}k}
\nonumber \\
& = & \prod_{n=1}^\infty \frac{(1-q^n)^2}{(1+ xq^{n-1/2})(1+ x^{-1}q^{n-1/2})}
\nonumber \\
& \stackrel{{\rm (by\,\, Eq. (\ref{F1}))}}{=\!\!=\!\!=\!\!=\!\!=\!\!=\!\!=\!\!=\!\!=\!\!=} & \Psi_A^{-1}(Y = x, X =q),
\\
\Psi_P(Y=-x, X= q)
& \stackrel{{\rm (by\,\, Eq. (\ref{W}))}}{=\!\!=\!\!=\!\!=\!\!=\!\!=\!\!=\!\!=\!\!=\!\!=} &
i(x+1)\Phi (x, q),
\\
\Psi_P(Y=-L, X=q) & \stackrel{{\rm (by\,\, Eq. (\ref{L}))}}{=\!\!=\!\!=\!\!=\!\!=\!\!=\!\!=\!\!=\!\!=\!\!=} &
i\sigma(L, q)\,.
\end{eqnarray}

\subsection{Affine (super)algebras, denominator identities}
\label{Affine}

For the affine Lie algebra $A_1^{(1)}$ computing the partition function $K(\beta)$
(see Sect. \ref{Weyl}), Eq. (\ref{K})) in low rank is equivalent to the following identity
\cite{Kac-Peterson}:
\begin{equation}
\prod_{n=0}^{\infty}(1-xq^n)^{-1}(1-x^{-1}q^{n+1})^{-1} =
\frac{\sum_{n\in {\mathbb Z}}(-1)^n(1-xq^n)^{-1}q^{(1/2)n(n+1)}}{\prod_{n=1}^\infty (1-q^n)^2}\,.
\label{EC1}
\end{equation}
Eq. (\ref{EC1}) then gives the denominator identity for the affine Lie algebra $A_1^{(1)}$ and also for the affine superalgebra ${\widehat{\mathfrak g}{\mathfrak l}}(1,1)$
\footnote{
The boson-fermion correspondence based on the Lie algebra $\widehat {{\mathfrak g}{\mathfrak l}}(1)$ lead to the denominator identity for $\widehat {{\mathfrak g}{\mathfrak l}}(2)$, whereas the super boson-fermion correspondence based on the Lie superalgebra $\widehat {{\mathfrak g}{\mathfrak l}}(1,1)$ produces the denominator identity for $\widehat {{\mathfrak g}{\mathfrak l}}(2,1)$. This parallel has been generalized for pairs of affine superalgebrtas $\widehat {{\mathfrak g}{\mathfrak l}}(m,n)$ and $\widehat {{\mathfrak g}{\mathfrak l}}(m+1,n)$ in \cite{KacW1}.}
\begin{equation}
\prod_{k= 1}^\infty \frac{(1-q^k)^2}{(1- xq^k)(1- x^{-1}q^{k+1})}
=
\frac{{\cR}^2 (s= 1-i\varrho(\tau))}
{{\cR}(s= (1- \overline{x})(1-i\varrho(\tau)))\cdot\cR(s= (2+ \overline{x})(1-i\varrho(\tau)))}\,.
\label{Id}
\end{equation}
Such identity can be obtained as a specialization of Ramanujan's summation formula for the bilateral
hypergeometric function ${}_1\psi_1$.
Recall that the general bilateral hypergeometric function ${}_r\psi_s(z)$ has the form \cite{Gasper}
\begin{equation}
{}_r\psi_s(z) =  \left[\begin{matrix}
a_1, a_2, \ldots, a_r \cr
b_1, b_2, \ldots, b_s
\end{matrix}; q,z\right]
= \sum_{n=-\infty}^{\infty}\frac{(a_1, a_2, \ldots, a_r; q)_n}{(b_1, b_2, \ldots; q)_n}
(-1)^{(s-r)n}q^{(s-r)n(n-1)/2}z^n,
\end{equation}
it assumes that each term of this serie is correctly defined.
The ${}_1\psi_1(a, b; q, zq)$-summation formula, which is due to Ramanujan, may be written as
\begin{equation}
{}_1\psi_1(a, b; q, zq) = \frac{(q, b/a, az, q/az; q)_\infty}{(b, q/a, z, b/az; q)_\infty}\,,
\,\,\,\,\,\,\, |b/a|< |z| < 1\,.
\end{equation}
Making use replacing $x\mapsto tq^{1/2}, \, {\rm log}\,x = \overline{x}{\rm log}\,q \mapsto
{\rm log}\,t = (\overline{t}-1/2){\rm log}\,q$ in Eq. (\ref{Id}) we obtain the denominator identity for the $N=2$ superconformal algebra (cf. Eqs. (\ref{Id2}) and (\ref{Id1})
\begin{eqnarray}
{\mathcal J}\left(\frac{1}{2}, \overline{t} -\frac{1+\tau}{2\tau}+\frac{1}{2}, 1,
\frac{1}{2}, 0; q\right)
& = &
\prod_{n= 1}^\infty \frac{(1-q^n)^2}{(1- tq^n)(1- t^{-1}q^n)}
\nonumber \\
& = &
(1-t)(1-t^{-1}q)\sum_{n\in {\mathbb Z}}(-1)^n\frac{q^{(1/2)n(n+1)}}{1-tq^n}\,.
\label{Id3}
\end{eqnarray}

\section{Concluding remarks}
\label{Concl}

In recent years $q$-series have attracted new interest in research areas such as topological field theory and Lie algebras, and have stimulated new developments in mathematical areas where $q$-series are familiar objects. The relationship between Lie algebras and combinatorial identities (the Euler identity, as one of its famous examples) was first discovered by Macdonald. A general procedure for proving combinatorial identities is based on the Euler-Poincar\'e formula.
Usually the Euler-Poincar\'e formula applies to chain complexes of finite dimensional Lie algebras.
In the infinite dimensional case matters can be fixed up by considering polygraded Lie algebras.

The partition functions can indeed be converted into product expressions; certain formulas for the partition functions (and Poincar\'{e} polynomials) are associated with dimensions of homologies of appropriate topological spaces and linked to generating functions and elliptic genera.
Note that this conclusively explains the sequence of dimensions ({\it distinguished powers}) of the simple Lie algebras (see Sect. \ref{Weyl}, Eq. (\ref{K})).

A set of combinatorial identities can be obtained by applying to subalgebras of Ka\v{c}-Moody algebras. Recall the construction of these subalgebras based on the Cartan matrix (see Sect. \ref{Weyl}). From the point of view of the applications, homologies associated with these
subalgebras constitute the thechnical basis of the proof of the combinatorial identities of
Euler-Gauss-Jacobi-MacDonald.

In various applications of conformal field theory, such as the Witten genus and its characteristic series, $KO$ characteristic series, we have shown that $q$-series can be expressed in a quite universal way in terms of classical special functions and Ruelle spectral functions related to the congruence subgroup $SL(2; {\mathbb Z})$. These mathematical tools allow us to derive interesting
new results (and important old results) in particular in superconformal field theory.

Given the significance of the Ruelle spectral function for the three dimensional hyperbolic geometry, this is a clear manifestation of the remarkable link that exists between all the above and hyperbolic three-geometry. More to the point, the link between two- and three-dimensional quantum models manifests itself in the fact that appropriate $q$-series (such as the characteristic series, the denominator identities, the characters of superconformal and affine Lie algebras) can be expressed
in a quite universal way in terms of spectral functions of hyperbolic geometry. Clarifying this connection is an intriguing challenge for the future.

\subsection*{Acknowledgments}

AAB would like to acknowledge the Conselho Nacional de Desenvolvimento
Cient\'ifico e Tecnol\'ogico (CNPq, Brazil)
and Coordena\c c\~{a}o de Aperfei\c coamento de Pessoal de N\'{\i}vel Superior (CAPES, Brazil) for financial support.
The support of the Academy of Finland under the Project No. 272919
is gratefully acknowledged.


\begin{thebibliography}{199}

\bibitem{Bytsenko13}
A. A. Bytsenko, M. Chaichian, R. J. Szabo and A. Tureanu, {\it Quantum Black Holes, Elliptic Genera
and Spectral Partition Functions},  Int. J. Geom. Meth. Mod. Phys. {\bf 11} (2014) 1450048 (42 pages);
[arXiv:hep-th/1308.2177].

\bibitem{Bonora14}
L. Bonora, A. A. Bytsenko and M. E. X. Guimarães, {\it Generalized q-Deformed Correlation Functions as Spectral
Functions of Hyperbolic Geometry}, Eur. Phys. J. C {\bf 74} (2014) 2976 (8 pages); [arXiv:hep-th/1405.4717].

\bibitem{Sharpe09}
M. Ando and E. Sharpe, {\it Elliptic genera of Landau-Ginzburg models over nontrivial spaces},
Nucl. Phys. B {\bf 414} (1994) 191-212; [arXiv:hep-th/9306096v2].

\bibitem{Bonora11}
L. Bonora and A. A. Bytsenko, {\it Partition functions for quantum gravity, black holes, elliptic genera
and Lie algebra homologies}, Nucl. Phys. B {\bf 852} (2011) 508-537; [arXiv:hep-th/1105.4571].

\bibitem{Witten87}
E. Witten, {\it Elliptic genera and quantum field theory}, Commun. Math. Phys. {\bf 109} (1987) 525-536.

\bibitem{Witten88}
E. Witten, {\it The index of the Dirac operator in loop space}, Elliptic Curves and Modular Forms in Algebraic Topology (New York. P.S. Landweber, ed.), Lecture Notes in Math. {\bf 1326} pp. 161-181, Springer-Verlag, 1988.




\bibitem{AndoFG}
M. Ando, C. P. French and N. Ganter, {\it The Jacobi Orientation and the Two-Variable Elliptic Genus},
Algebraic and Geometric Topology {\bf 8} (2008) 493-539; [arXiv:math/0605554].

\bibitem{Gosper}
W. Gosper, {\it A Calculus of Series Rearrangements, Algorithms and Complexity} (J. F. Traub Ed.),
pp. 121-151, Academic Press, 1976.

\bibitem{Askey}
R. Askey, {\it Ramanujun's extensions of the gamma and beta functions}, Amer. Math. Monthly
{\bf 87} (1980) 346-359.

\bibitem{KacW}
V. G. Ka\v{c} and M. Wakimoto, {\it Integrable highest weight modules over affine superalgebras and number theory}, Progress in Math. {\bf 123} (1994) 415-456; [arXiv:hep-th/9407057].

\bibitem{Kac-Peterson}
V. G. Ka\v{c} and D. H. Peterson, {\it Infinite-Dimensional Lie Algebras, Theta Functions and Modular Forms}, Advances Math. {\bf 53} (1984) 125-264.

\bibitem{Hecke}
E. Hecke, {\it \"{U}ber den Zusammenhang zwischen elliptischen Modulfunktionen und indefiniten quadratischen Formen}, Math. Werke, Vandenhoeck and Ruprecht, pp. 418-427, G\"{o}ttingen, 1959.

\bibitem{Rogers}
L. J. Rogers, {\it On the expansion of some infinite products}, Proc. Londion Math. Soc. {\bf 24}
(1893) 337-352.

\bibitem{Kac80}
V. G. Ka\v{c} and D. H. Peterson, {\it Affine Lie algebras and Hecke modular forms},
Bull. Amer. Math. Soc. {\bf 3} (1980) 1057-1061.

\bibitem{Cherednik}
I. Cherednik and B. Feigin, {\it Rogers-Ramanujan type identities and Nil-DAHA},
Advances Math. {\bf 248} (2013) 1059-1088; [arXiv:math.QA/1209.1978v4].

\bibitem{Bressound}
D. M. Bressound, {\it Hecke modular forms and $q$-Hermite polynomials}, Illinois J. Math.
{\bf 30} (1986) 185-196.

\bibitem{Andrews}
G. E. Andrews, {\it q-Series: Their Development and Application in Analysis, Number Theory, Combinatorics,
Physics, and Computer Algebra}, Expository Lectures from the CBMS Regional Conference No. 66, Providence, Rh. I.: AMS, 1986.

\bibitem{Matsuo}
Y. Matsuo, {\it Character formula of $c< 1$ unitary representation of $N=2$ superconformal algebra},
Progr. Theor. Phys. {\bf 77} (1987) 793-797.

\bibitem{Dobrev}
V. K. Dobrev, {\it Characters of the unitarizable highest weight modules over the $N=2$ superconformal algebras}, Phys. Lett. B {\bf 186} (1987) 43-51; [arXiv:hep-th/0708.1719].

\bibitem{Kiritsis}
E. B. Kiritsis, {\it Character formulae and the structure of the representations of the $N=1,\, N=2$
superconformalal algebras}, Int. Jour. Mod. Phys. A {\bf 3} (1988) 1871 (38 pages).



\bibitem{KacW1}
V. G. Ka\v{c} and M. Wakimoto, {\it Integrable highest weight modules over affine superalgwebras and
Appell's function}, Commun. Math. Phys. {\bf 215} (2001) 631-682; [arXiv:math-ph/0006007v1].

\bibitem{Gasper}
G. Gasper and M. Rahman, {\it Basic Hypergeometric Series}, Encyclopedia of Mathematics and its Applications, Ed. by G.-C. Rota, {\bf 35} Cambridge University Press 1990.



\end{thebibliography}
\end{document}